\documentclass[prb,final,twocolumn,superscriptaddress]{revtex4}
\usepackage{graphicx}

\begin{document}

\title{Unconventional Hall effect in oriented Ca$_3$Co$_4$O$_9$ thin films}

\author{H.W. Eng}
\affiliation{Laboratoire CRISMAT, UMR 6508 CNRS, ENSICAEN, 6, Bd du Mar\'echal Juin, 14050 CAEN Cedex, France}
\author{P. Limelette}
\thanks{Present address: Laboratoire LEMA, UMR 6517 CNRS-CEA, Universit\'e Rabelais, UFR Sciences, Parc de Grandmont, 37200 Tours, France}
\affiliation{Groupe Mati\`ere Condens\'ee et Mat\'eriaux, UMR CNRS 6626, Universit\'e de Rennes I, 35042 Rennes,  France}
\author{W. Prellier}
\thanks{prellier@ensicaen.fr}
\affiliation{Laboratoire CRISMAT, UMR 6508 CNRS, ENSICAEN, 6, Bd du Mar\'echal Juin, 14050 CAEN Cedex, France}
\author{Ch. Simon}
\affiliation{Laboratoire CRISMAT, UMR 6508 CNRS, ENSICAEN, 6, Bd du Mar\'echal Juin, 14050 CAEN Cedex, France}
\author{R. Fr\'esard}
\affiliation{Laboratoire CRISMAT, UMR 6508 CNRS, ENSICAEN, 6, Bd du Mar\'echal Juin, 14050 CAEN Cedex, France}

\begin{abstract}
Transport properties of the good thermoelectric misfit oxide Ca$_3$Co$_4$O$_9$ are examined. 
In-plane resistivity and Hall resistance measurements were made on epitaxial thin films which were grown on {\it c}-cut sapphire substrates using the pulsed
laser deposition technique. Interpretation of the in-plane transport experiments relates the substrate-induced strain in the resulting film to
single crystals under very high pressure ($\sim$ 5.5 GPa) consistent with a key role of strong electronic correlation. 
They are confirmed by the measured high temperature maxima in both resistivity and Hall resistance. 
While hole-like charge carriers are inferred from the Hall effect measurements over the whole investigated temperature range, 
the Hall resistance reveals a non monotonic behavior at low temperatures that could be interpreted with an anomalous contribution. 
The resulting unconventional temperature dependence of the Hall resistance seems thus to combine high temperature strongly correlated 
features above 340 K and anomalous Hall effect at low temperature, below 100 K.
\end{abstract}

\maketitle
Good thermoelectric materials,\cite{Mahan,DiSalvo} which convert heat into electricity and vice versa, should have high figures of merit (ZT) where 
ZT= S$^2$T/$\rho \kappa $, so S (the thermoelectric power or Seebeck coefficient) should be large while $\rho $ (resistivity) and $\kappa $ 
(thermal conductivity) should be small at a temperature T. 
In addition to these properties, these materials should be physically and chemically robust for high temperature processes such as the generation of
energy from waste heat, and therefore, oxides such as Na$_x$CoO$_2$
and Ca$_3$Co$_4$O$_9$ have been received a considerable attention
recently.\cite{Terasaki,Masset}  
Among the oxides Ca$_3$Co$_4$O$_9$ is very promising because of its high room temperature (RT) thermopower {\bf (}125 $\mu $V/K {\bf ),} low
resistivity (12 m$\Omega$ cm), low thermal conductivity (30 mW (cm K)$^{-1}$), and resistance to humidity.\cite{Masset,Satake}

Ca$_3$Co$_4$O$_9$ is a misfit oxide and can be denoted as [Ca$_2$CoO$_3$]$^{RS}$[CoO$_2$]$_{1.62}$ to recognize the incommensurate nature of
the structure. The structure is composed of alternating layers of a distorted Ca$_2$CoO$_3$ rock salt-like layer (RS) and a CoO$_2$ cadmium
iodide-like layer which are stacked in the {\it c}-axis direction.
Crystallographically, these two layers have similar {\it a,} {\it c}, and {\it $\beta$} lattice parameters but different {\it b} lattice
parameters. The ratio of the {\it b} parameters for the Ca$_2$CoO$_3$ layer to CoO$_2$ layer is 1.62. The material's anisotropic behavior is
easily seen through the comparison of the in-plane and out-of-plane resistivity behavior (metallic versus semiconducting, respectively).\cite{Masset} 
Thus, to attain the highest properties in a thermoelectric device, growth along the {\it c}-axis would be best to insure uniform properties. The bulk can be
magnetically aligned along the {\it c}-axis at high temperature,\cite{Sano,Horii} but the thin films form provides a more convenient method for
manufacturing useful thermoelectric devices. 
For this reason, we have undertaken the synthesis of Ca$_3$Co$_4$O$_9$ films on Al$_2$O$_3$ ($c$-cut sapphire) substrates.\cite{Eng} 
The resulting epitaxial film has thermoelectric properties very close to the single crystal (110 $\mu$V K$^{-1}$ at RT)
but with slightly higher resistivity values (varying from 11 to 21 m$\Omega$ cm).

The underlying theory behind the good thermoelectric values for Ca$_3$Co$_4$O$_9$ is still not completely understood despite the electronic structures having
 been studied,\cite{Asahi} thus more experimental data is needed. From a fundamental aspect, the good properties of the cobaltites (Ca$_3$Co$_4$O$_9$ and
 Na$_x$CoO$_2$) are not expected because conventional theory states that carrier concentration should oppositely affect conductivity and
thermoelectric power, e.g., high carrier concentrations should lead to high conductivity and low thermoelectric power.\cite{Mahan} 
High-resolution photoemission spectroscopy has been used to demonstrate that the high thermoelectric power and low conductivity depend upon a narrow conduction
 band derived from the two dimensional CoO$_2$ layers in the structures of Ca$_3$Co$_4$O$_9$, Bi$_2$Sr$_2$Co$_2$O$_9$, and Na$_x$CoO$_2$.\cite{Takeuchi}
Furthermore, transport properties in Ca$_3$Co$_4$O$_9$ single crystals were found to display typical features of strongly correlated 
electron materials.\cite{Limelette}

Viewing these results, it is interesting to study the transport and Hall effect measurements of Ca$_3$Co$_4$O$_9$ made in the form of thin films, and our
results are reported in this paper. We are studying the properties of thin films for two reasons:

(1) the resulting thin film is strongly oriented in contrast to bulk ceramic, allowing a clear assessment of the anisotropic properties.

(2) thin films allow a thorough and rapid examination of the transport properties, in particular the Hall effect properties which can give
important information on the conduction mechanisms and the nature of the carriers. 
Unfortunately Hall effect experiments are generically difficult with single crystals but can be much more easily performed with thin films.

Thin films of Ca$_3$Co$_4$O$_9$ were grown on polished $c$-cut ({\it 0001}-cut) sapphire substrates using pulsed-laser deposition technique.\cite{PLD} A black,
 sintered Ca$_3$Co$_4$O$_9$ target was synthesized using conventional solid state techniques,\cite{Masset} and a KrF excimer laser beam (Lambda Physik
 Compex, $\lambda $=248 nm, repetition rate=3 Hz) was used to deposit the material composition of the target onto the sapphire substrate at the proper
 deposition conditions (600 ${{}^{\circ }}$C, 0.1 mbar O$_2$ pressure, and a fluence of 1.7 J cm$^{-2}$). 
For more details of the optimization and growth conditions see Ref.\cite{Eng} 
The x-ray diffraction of the resulting films, attesting to their good quality and good epitaxial relation with the substrate, was shown previously.\cite{Eng} 
A DekTak$^3$ST surface profiler found the thickness of the thin films to be 1800 \AA .

\begin{figure}[htbp]
\centerline{\includegraphics[width=0.95\hsize]{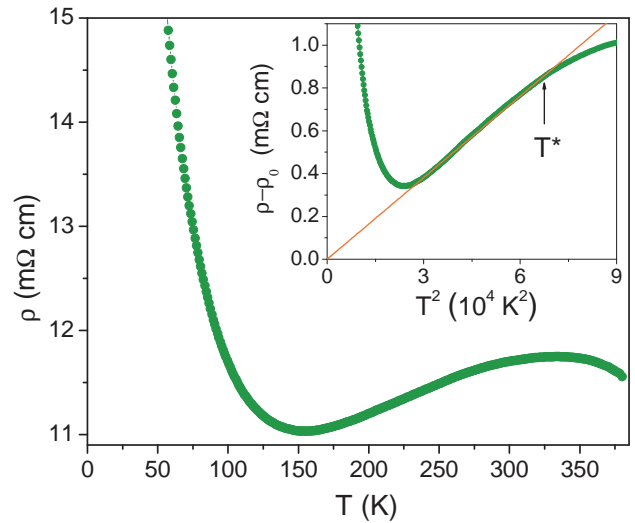}}
\caption{(Color online) Longitudinal resistivity of a representative Ca$_3$Co$_4$O$_9$ thin film. 
The inset displays the temperature region where $\rho -\rho _0$ versus T$^2$ is linear. 
T$^{*}$ notes a cross-over above which the resistivity no longer varies as T$^2$.
}
\label{fig1}
\end{figure}

The transport measurements were performed using a Quantum Design Physical Property Measurement System (PPMS). The standard four point probe method was used
 to measure resistivity, or more specifically, longitudinal resistivity ($\rho _{xx}$). To make the appropriate connections onto the film, four silver plots
 were first deposited via thermal evaporation onto the film, and then thin aluminum contact wires were used to connect the areas to the electrodes.
For Hall effect measurements, a silver layer and then a gold layer were deposited onto the surface of the film before standard ultraviolet
photolithography and argon ion etching were used to pattern microbridges (where the largest widths were 100 $\mu $m). The silver layer was deposited
via thermal evaporation, and the gold layer was deposited via the rf sputtering technique.
The resistivity results, shown in the Fig.~\ref{fig1}, are similar in behavior to the $c$-axis aligned transport properties of the bulk.\cite{Sano,Horii} 
There are three interesting regions to note in the Fig.~\ref{fig1}: 
below 160 K where the resistivity increases monotonically as temperature T decreases, between 160 and 340 K where the resistivity is metallic-like, 
and above 340 K where the resistivity decreases as T increases.

It is appealing to analyze the low temperature resistivity using an activation behavior, with an activation temperature T$_0$ since, below T$_{inc}$=100K,
 muon spin spectroscopy experiments showed the existence of a short-range incommensurate magnetic order in bulk samples.\cite{sugiyama} 
For sample \#1, we find T$_0$=13 K. Since $T_0 \ll T_{inc}$, our low temperature resistivity data  
cannot be interpreted in the framework of an activation law, and they require further investigations.

Additionally, two noticeable differences from the bulk are (a) higher resistivity values and (b) a more shallow minima. A possible reason for the increased
 resistivity can be an increase in grain boundaries in the film, as is commonly seen in many thin metal films.\cite{Kasap} Previous bulk studies 
\cite{Sano,Karpinnen} have found that oxygen deficiency increased both the resistivity and the thermoelectric power. But oxygen defects are unlikely to be the
predominant source of the higher resistivity values because our films showed only an increase in resistivity and not in thermoelectric power.
\begin{table}[b!]
\caption 
{The residual resistivities ($\rho _0$), Fermi liquid transport coefficients (A), and the transport cross-over temperatures (T$^{*}$) 
for three thin films made at similar deposition conditions (600 ${{}^{\circ }}$C, 0.1 mbar O$_2$ pressure, and a fluence of 1.7 J cm$^{-2}$) on $c$-cut
 sapphire substrates.
}
\vskip .2cm
\begin{ruledtabular}
\begin{tabular}{cccc}
\multicolumn{1}{c}{Sample}           &
\multicolumn{1}{c}{$\rho_0$ (m$\Omega$ cm)}           &
\multicolumn{1}{c}{A (10$^{-5}$ m$\Omega$ cm K$^{-2}$)}           &
\multicolumn{1}{c}{T$^*$ (K)}           \cr
\colrule
1  & 10.65  & 1.27 & 240     \\
2  & 10.69  & 1.02 & 231     \\ 
3  & 20.93  & 1.14 & 240    \\
\end{tabular}
\end{ruledtabular}
\label{tab:egF}
\end{table}

In bulk samples, a small anomaly of the resistivity has been observed
at 380K,\cite{Masset,Limelette,sugiyama} and interpreted as a spin
state transition.\cite{Sugiyama_03} It is not observed here. 

In agreement with a previous paper devoted to pressure effects in bulk samples \cite{Limelette} the metallic portion of the transport curve 
is consistent with a Fermi liquid regime, from nearly 160 K up to a characteristic temperature T$^* \sim$ 240 K, with a resistivity varying as 
$\rho$=$\rho _0$+AT$^2$. 
The first term, $\rho _0$, is the residual resistivity and depends on extrinsic factors such as scattering at the grain boundaries.
As can be seen in Table~\ref{tab:egF}, the values of $\rho _0$ vary even under similar growth conditions. 
The second term depends on intrinsic effects and is representative of the strength of the electronic correlations. 
Strongly enhanced in the vicinity of a Mott Metal-Insulator
transition,\cite{DMFT} the high values of the A coefficients listed in
Table~\ref{tab:egF} (from 1.02 to 
 1.27x10$^{-5}$ m$\Omega$ cm K$^{-2}$) are rather sample independent and imply strong interactions between electrons.
\begin{figure}[htbp]
\centerline{\includegraphics[width=0.95\hsize]{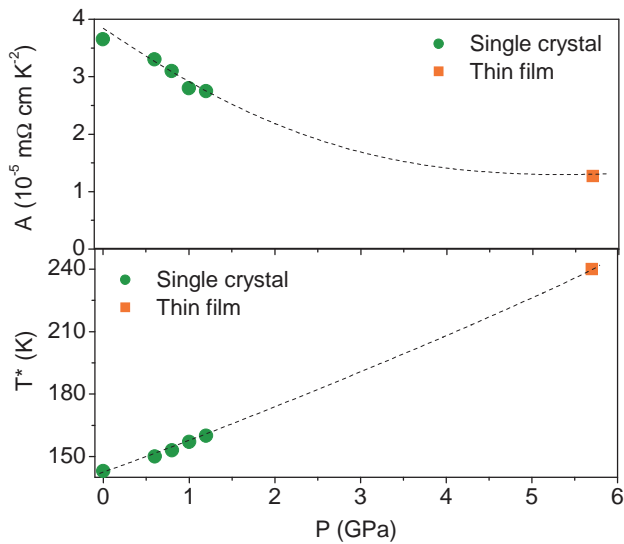}}
\caption{(Color online) Single crystal (circles) and extrapolated thin film (squares) values for the coefficients of the metal-like resistivity as a
 function of pressure: A, the Fermi liquid transport coefficient (upper panel) and T$^{*}$, the transport cross-over temperature (lower panel). 
The strain induced pressure was extrapolated from the values of T$^{*}$. The two dotted lines are guides to the eyes.
}
\label{fig2}
\end{figure}

By comparing in Fig.~\ref{fig2} the pressure dependences of single crystal transport properties \cite{Limelette} with the deduced ones for the studied
thin film, a lower value of the coefficient A is observed in the thin films suggesting here lower correlations. 
The transport cross-over temperature T$^{*}\sim$ 240 K that ends the quadratic Fermi liquid regime is found to be higher than in single 
crystals in agreement with a lower coefficient A. 
Consistently, one must emphasize that the product A(T$^*)^2$/b$_2$, with b$_2$ the in-plane lattice parameter in the  CoO$_2$ layer,\cite{Masset} is of 
the order of h/e$^2$ as expected in a Fermi liquid.\cite{DMFT} Indeed, the Fermi liquid transport coefficient A is proportional to the square of 
the electronic effective mass $m^*$ while the transport cross-over temperature T$^{*}$ represents an effective Fermi temperature, namely renormalized 
by correlations as T$^{*}$=T$_F$/m$^*$. Thus, the product A(T$^*)^2$ remains independent of the electronic correlations as observed experimentally in both
 thin films and single crystals by varying pressure.\cite{Limelette}

From an extrapolation of the single crystal data, the film's T$^{*}$ yields an approximate strain induced pressure of approximately 5.5 GPa (Fig.~\ref{fig2}).
 This relatively high pressure extrapolation is not surprising since the in plane structural parameter of the film, assumed to be equal to the parameter of
 the substrate in such epitaxial films, is smaller by a few percents than the parameter of the single crystal and roughly corresponds to such a pressure. 

This strain effect makes possible to grow as a thin film metastable phases (e.g., infinite layers \cite{Il} and oxycarbonate superconductors 
\cite{Oxycarbonate}) as those synthesized using the high pressure processes.\cite{HP1,HP2} 
It should be pointed out here that the induced pressure is biaxial, in contrast to what is obtained in the case of hydrostatic pressure already published.
 In addition, it is well known in these systems that there exists a relaxation of the strains along the thickness which can also broadens the whole result.
\begin{figure}[t!]
\centerline{\includegraphics[width=0.95\hsize]{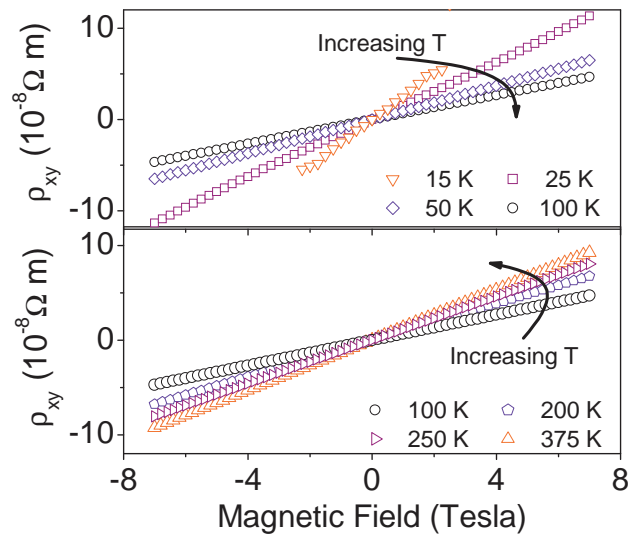}}
\caption{(Color online) Hall effect measurements of the Ca$_3$Co$_4$O$_9$ thin films. The transverse resistivity ($\rho _{xy}$) versus magnetic field is
 shown for values above 100 K (upper panel) and below 100 K (lower panel).
}
\label{fig3}
\end{figure}

\begin{figure}[b!]
\centerline{\includegraphics[width=0.95\hsize]{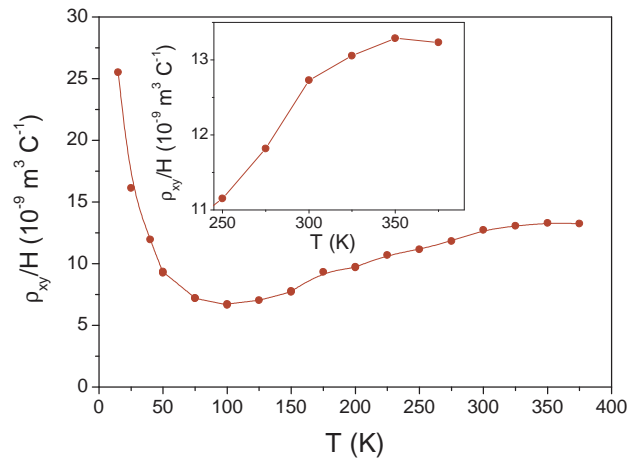}}
\caption{(Color online) Temperature dependence of the Hall resistance ($\rho_{xy}$/H). 
A broad maximum is observed at high temperatures (inset).
}
\label{fig4}
\end{figure}

Hall effect measurements of the thin films are shown in Figure 3. 
(The Hall resistance is equal to the transverse resistivity divided by the magnetic field ($\rho _{xy}$/H) which is related to the applied magnetic field H.) 
First of all, let us mention that the positive slope of the Hall resistance implies hole-like charge carriers 
as inferred from thermoelectic power measurements.\cite{Masset}
Moreover, while in a regular metal the Hall resistance would be nearly temperature independent, the Fig.~\ref{fig4} exhibits 
an unconventional temperature dependence with a highly non monotonic behavior. 
In particular, one observes a sizeable increase from 100 K up to 350 K followed by a broad maximum at high temperatures displayed in the inset 
of the Fig.~\ref{fig4}. It is inconsistent with the linear increase of the Hall coefficient predicted in the charge frustration scenario.\cite{Motrunich} 

Expected within the framework of the Dynamical Mean-Field Theory (DMFT) of strongly correlated systems,\cite{DMFT} the maxima 
in both resistivity and Hall resistance result from a strong temperature dependence of the density of states leading to a pseudo gap  above T$^{*}$.
Let us mention that this unusual behavior has already been observed in the strongly correlated quasi-2D organic superconductors, [$\beta$-(BEDT-TTF)$_2$I$_3$]
.\cite{KHFC}
In addition, even assuming a constant value in this range, the classical analysis drives to unrealistic results. 
Indeed, the corresponding number of carriers n is estimated to be 5 10$^{26}$ carriers m$^{-3}$ by the inverse relationship to the Hall effect coefficient
 which is used for common metals (R$_H$=1/ne). Thus, these results firmly suggest that the observed strong temperature dependence of the Hall resistance 
including the broad maximum originates from the strong electronic correlations.
\begin{figure}[t!]
\centerline{\includegraphics[width=0.95\hsize]{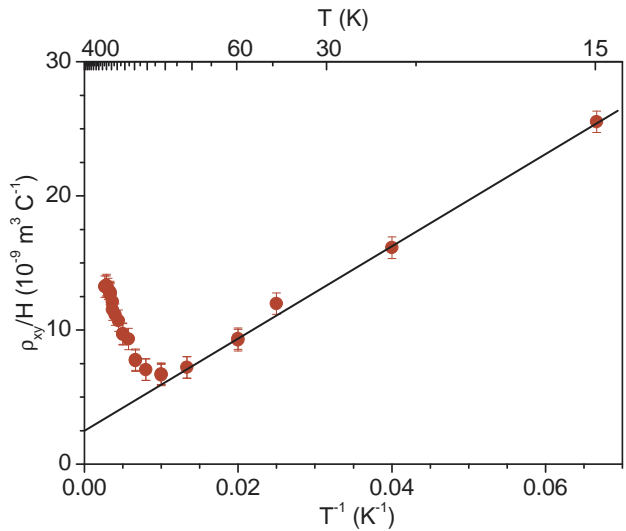}}
\caption{(Color online) The Hall resistance ($\rho _{xy}$/H) values of the thin film versus inverse temperature. 
A linear relationship is observed below 100 K as $\rho _{xy}$/H=2.6 10$^{-9}$(1+132/T).
}
\label{fig5}
\end{figure}

Let us now discuss the Hall resistance increase at low temperature. 
As displayed in Fig.~\ref{fig5}, $\rho _{xy}$/H varies linearly with the inverse temperature below 100 K as $\rho _{xy}$/H=2.6 10$^{-9}$(1+132/T).
In order to analyze this behavior, the latter relation must be compared to the expression $\rho _{xy}$/H=R$_H$+((1-N)R$_H$+R$_S$)M/H where N is the shape
 of the sample, R$_H$ the ordinary Hall coefficient, R$_S$ the anomalous Hall coefficient \cite{Hurd} which originates from spin-orbit effects, and M the
 magnetization.\cite{lut,noz}

Since unfortunately we were not able to measure the magnetization in the film because of the high contribution of the substrate, let us consider the bulk
 one.

This can be estimated to be of Curie type with a Curie constant $C_0$ of the order of 0.12 K,\cite{Masset} that should be related to the transverse 
resistivity following Eq.~\ref{eq}. 
\begin{equation}
\rho _{xy}/H \approx R_H (1+(1+R_S/R_H) C_0/T) 
\label{eq}
\end{equation}

Thus, by comparing the experimental relation with Eq.~\ref{eq}, one deduces a strong anomalous component as R$_S$/R$_H > $ 10$^3$.
Even if a quantitative analysis of R$_S$ would require magnetization measurements in the films as a function of temperature, 
its order of magnitude compared to R$_H$ testifies unambiguously to this anomalous component.\\

Furthermore, since the ordinary Hall coefficient is expected to be weakly temperature dependent at low temperature, the experimental relation (of Eq. 1)
 gives also an estimate of R$_{H} \approx$ 2.6 10$^{-9}$ m$^{3}$ C$^{-1}$ (see Fig.5).
Because in this limit the ordinary Hall coefficient is independent of the electronic correlations,\cite{DMFT} this value leads to a reasonable hole-like 
charge carriers density n $\approx$ 2.4 10$^{+27}$ m$^{-3}$ with R$_H$=1/ne. 
Interestingly, the latter density corresponds to a unit cell volume V $\approx$ 1.48 10$^{-28}$ m$^{-3}$ to a hole doping of the order of 0.36 
per cobalt of the CoO$_2$ layer. 
Assuming an insulating rock salt-like layer, this doping is consistent with a crude estimation inferred from the electroneutrality in the formula 
[Ca$_2$CoO$_3$]$^{RS}_{0.62}$[CoO$_2$]. 
Thus, it could suggest that the transport properties are essentially governed by itinerant holes in the CoO$_2$ layers.

In summary the in-plane resistivity and Hall effect measurements made on epitaxial thin films of Ca$_3$Co$_4$O$_9$ grown on 
$c$-cut \hspace{0pt}sapphire substrates confirmed the strong electron correlations. In the thin film form, the in-plane transport properties
verified that the strain from the substrate can be related to very high pressures ($\sim$ 5.5 GPa) in single crystals. 
By using the strain induced by the substrate as an equivalent pressure, we have shown that the effective mass of the charge carriers can be 
significantly modified. 
The strong electron correlation nature is borne out in the high temperature Hall effect measurements where the observed Hall resistance plateaus and
 no longer follows expected Fermi liquid behavior. 
Finally at low temperatures, the observed high Hall resistances may be attributed to an anomalous Hall effect and should be related to
 the magnetic susceptibility as well as the appearance of the short-range IC-SDW order.The misfit nature of Ca$_3$Co$_4$O$_9$ effectively localizes 
the itinerant holes in the CoO$_2$ layers, leading to a narrow conduction band that yields strongly correlated transport properties.

\begin{acknowledgments}
The funding for this project was provided by the Centre National de la
Recherche Scientifique (CNRS) and the Conseil R\'{e}gional de
Basse-Normandie. We would also like to thank Dr. L. M\'{e}chin for
patterning the microbridges on the thin films and Prof. B. Mercey,
Dr. M.~Singh, and Dr. P. Padhan for helpful discussions.
\end{acknowledgments}

\end{document}